\documentclass[epj]{svjour}
\usepackage{graphics}
\usepackage{amsmath}
\usepackage{placeins}
\usepackage{hyperref}
\usepackage{amssymb,bm,calc}
\begin{document}
%

\title{Final-state effects on the transverse-momentum spectra of charged
hadrons in $p+Pb$ and $Pb+Pb$ collisions at $\sqrt {s_{\rm{NN}}}~=~5.02$~TeV 
using the modified Tsallis distribution}

\author{Kapil Saraswat\inst{1}$^{\ast}$ \and 
Deependra Singh Rawat\inst{2}          \and
Manoj Kumar Singh\inst{3}$^{\dagger}$   \and
Prashanta Kumar Khandai\inst{4}        \and
Damini Singh\inst{5}                   \and
Venktesh Singh\inst{6}
}
\institute{
Department of Physics, School of Sciences, Noida International University,
Plot No. 1, Sector 17A, Yamuna Expressway, Greater Noida,
Gautam Buddha Nagar - 203201, India.
\and
School of Allied Sciences (Physics), Graphic Era Hill University,
Bhimtal Campus, Sattal Road, Nainital - 263136, India.
\and
Institute of Physics, Academia Sinica, Taipei - 115201, Taiwan.
\and
Department of Physics, Ewing Christain College, Prayagraj - 212003, India.
\and
Department of Physics, Gujarat Arts and Science College, Ellisbridge,
Ahmedabad - 380006, India.
\and
Department of Physics, School of Physical and Chemical Sciences, 
Central University of South Bihar,
Gaya - 824236, India.}

\abstract{
The ATLAS Collaboration has reported measurements of the transverse
momentum ($p_{\rm{T}}$) spectra of charged hadrons in proton-proton ($p+p$),
proton-lead ($p+Pb$), and lead-lead ($Pb+Pb$) collisions at a
nucleon-nucleon center-of-mass energy of $\sqrt{s_{\rm{NN}}}~=$~5.02~TeV
within the rapidity interval $-2.5<y<2.0$. In the present work, we
investigate medium effects on charged-hadron production in $p+Pb$ and
$Pb+Pb$ collisions using a phenomenological modified Tsallis
parametrization that accounts for transverse collective flow in the
low-to-intermediate $p_{\rm{T}}$ region and medium-induced parton energy
loss at high $p_{\rm{T}}$. The modified parametrization provides a
consistent description of the measured spectra over the full investigated
$p_{\rm{T}}$ range across different centrality classes. The extracted
energy-loss exponent $\alpha$, which characterizes the energy dependence
of parton energy loss, is found to lie in the range 0.59$-$0.73 for
$p+Pb$ collisions and 0.32$-$0.59 for $Pb+Pb$ collisions.
In addition, the extracted parameters exhibit a clear centrality
dependence, reflecting enhanced collective effects in central collisions.
These results provide quantitative insight into the interplay between
transverse collective flow and medium-induced parton energy loss in
charged-hadron production at LHC energies.  
}
\maketitle
\footnote{$^{\ast}$ E-mail : drkapilsaraswat@zohomail.in}
\footnote{$^{\dagger}$ man.bhu9@gmail.com}

\section{Introduction}
The RHIC (Relativistic Heavy Ion Collider) \cite {RHIC1}\cite{RHIC2} and
the LHC (Large Hadron Collider)~\cite{LHC} are employed to investigate a
state of matter known as the quark-gluon plasma (QGP) \cite{Shuryak}.
The QGP is a deconfined, strongly interacting state of quarks and gluons
in thermal equilibrium. Such a state is believed to have existed in the
early Universe shortly after the Big Bang \cite{Olive} \cite{Schwarz}.
A variety of heavy-ion collision systems have been studied at RHIC,
including $Au+Au$, $Cu+Cu$, and $Cu+Au$, and at the LHC, including
$Pb+Pb$ and $Xe+Xe$. Proton-proton ($p+p$) collision serves as a
baseline for heavy-ion ($A+A$) collisions \cite{PPPROD}.
Particle production in both $p+p$ and $A+A$ collisions arises from
multiple partonic scatterings. The transverse momentum ($p_{\rm{T}}$)
spectra in $p+p$ collisions provide insight into the freeze-out stage,
where particles cease to interact. Recent high-multiplicity $p+p$ data
from the LHC \cite{HighMul} \cite{HighMul2} suggest evidence for the
formation of a QGP-like medium.

Heavy-ion collisions at the LHC and RHIC provide the experimental
facilities for exploring the QCD phase diagram in both its perturbative
and non-perturbative regimes \cite{Shuryak} \cite{Cleymans:1985wb} 
\cite{Singh:1992sp} \cite{Jaiswal:2020hvk} \cite{Saraswat:2022zcn} 
\cite{Cho:1979nv} \cite{Cho:2002iv} \cite{Chandola:2009zz}.
In high-energy collisions of heavy nuclei such as lead or gold,
the momentum distributions of produced hadrons exhibit medium-induced
effects arising from the hot and dense strongly interacting matter
formed in the initial stage. These effects include collective
flow \cite{Collective_flow}, driven by the hydrodynamic expansion
of the medium, and jet quenching \cite{Jet_Quenching}, in which
high-energy partons lose energy while traversing the medium.

In $p+p$ collisions, the $p_{\rm{T}}$ spectra of hadrons are
commonly described using a Tsallis distribution
 \cite{Tsallis:1987eu} \cite{Biro:2008hz} characterized
by two parameters, $T$ and $q$ \cite{PPG099}. The parameter
$T$ is often interpreted as
the kinetic freeze-out temperature in heavy-ion collisions,
where inelastic and elastic interactions cease, whereas its
physical interpretation in $p+p$
collisions remains less well defined. The parameter $q$,
known as the nonextensivity parameter, quantifies deviations
from complete thermal equilibrium \cite{q_Tsallis}.
The Tsallis distribution, which effectively describes systems
close to thermal equilibrium, exhibits a functional form similar
to the ``Hagedorn power-law'' parametrization used to model
hard-scattering processes in quantum chromodynamics (QCD) 
\cite{IJMPA_Khandai} \cite{Wong:2012zr} \cite{Wong:2013sca}
 \cite{Hagedorn:1983wk} \cite{Blankenbecler:1974tm}.

The primary motivation of this work is to investigate the
impact of the hot and dense QGP, produced in relativistic
heavy-ion collisions, on the $p_{\rm{T}}$ spectra of charged hadrons.
While the Tsallis distribution successfully describes the $p_{\rm{T}}$
spectra in $p+p$ collisions, it does not fully reproduce the spectra
in $p+Pb$ and $Pb+Pb$ collisions, particularly in the high-$p_{\rm{T}}$
region ($p_{\rm{T}}~\gtrsim~7$~GeV/c), where final-state effects such as
collective flow and partonic energy loss
in the medium become significant. This motivates the development of a
phenomenological model capable of simultaneously describing both
low-$p_{\rm{T}}$ collective behavior and high-$p_{\rm{T}}$ suppression
across the full momentum range. In the current work, we employ a
modified Tsallis parametrization \cite{JPC_Kapil} \cite{Khandai} to
analyze the latest available charged-hadron $p_{\rm{T}}$ spectra
reported by the ATLAS Collaboration~\cite{ATLAS:2022kqu} for $p+Pb$
and $Pb+Pb$ collisions at a nucleon-nucleon center-of-mass energy of
$\sqrt{s_{\rm{NN}}}~=$~5.02~ TeV within the rapidity interval
$-2.5 < y < 2.0$.
The analysis is performed for different centrality classes, with both
statistical and systematic uncertainties incorporated in the fitting
procedure.
The extracted parameters are subsequently used
to investigate the centrality dependence of transverse collective flow
and medium-induced parton energy loss, providing quantitative insight
into the medium effects governing charged-hadron production
in relativistic heavy-ion collisions.
\FloatBarrier

\section{Conventional and modified Tsallis parametrizations}
The thermodynamically motivated Tsallis distribution \cite{Tsallis:1987eu}
\cite{Biro:2008hz} is given by
\begin{eqnarray}
E\frac{d^{3}N}{dp^{3}}~{=}~
C_{n}m_{T}\bigg(1+(q-1)\frac{m_{T}}{T}\bigg)^{-\left(\frac{1}{q-1}\right)}~,
\label{Tsallis}
\end{eqnarray}
where $C_{n}$ denotes the normalization constant, while $T$
and $q$ are the Tsallis parameters. The parameter $T$ is
commonly interpreted as the effective temperature, whereas 
$q$ quantifies the degree of nonextensivity of the system.
Here, $m_{T}~{=}~\sqrt{p_{\rm T}^{2} + m^{2}}$ represents the
transverse mass, where $m$ is the rest mass of the produced
hadron, and the power-law exponent is defined as
$n~{=}~\left(\frac{1}{q-1}\right)$.
This distribution successfully describes the $p_{\rm T}$ spectra
of hadrons produced in $p+p$ collisions over a broad
$p_{\rm T}$ range.

To account for final-state medium effects in heavy-ion
collisions, a modified Tsallis distribution \cite{JPC_Kapil},
incorporating transverse collective flow \cite{Tang:2008ud}
\cite{Khandai:2013fwa} \cite{Sett:2015lja} in the
low-$p_{\rm T}$ region and in-medium energy loss in the
high-$p_{\rm T}$ region, is employed:

\begin{subequations}
\label{modified_new_func_tsallis_distribution_function}
\begin{align}
\label{new_func_tsallis_distribution_function}
E\frac{d^{3}N}{dp^{3}} &= A_{1}\bigg[{\rm exp}\bigg(-\frac{\beta p_{\rm T}}{p_{1}}\bigg) + \frac{m_{T}}{p_{1}}\bigg]^{-n_{1}}~, ~~ p_{\rm T}<p_{\rm T}^{th}~, \\
\label{new_func_tsallis_distribution_function_second}
E\frac{d^{3}N}{dp^{3}} &= A_{2}\bigg[\frac{B}{p_{2}}\bigg(\frac{p_{\rm T}}{q_{0}}\bigg)^{\alpha} + \frac{m_{T}}{p_{2}}\bigg]^{-n_{2}}~, ~~~~~~~~ p_{\rm T}>p_{\rm T}^{th}~, 
\end{align} 
\end{subequations}
where Eq.~\ref{new_func_tsallis_distribution_function}
describes the low-$p_{\rm T}$ component of the hadron
spectrum, incorporating thermal production and transverse
collective effects. The parameter $A_{1}$ represents the
normalization factor of this component, while $n_{1}$
controls its spectral shape. The effective temperature
is defined as $T~{=}\left(\frac{p_{1}}{n_{1}}\right)$,
where $p_{1}$ is the characteristic momentum scale of the
low-$p_{\rm T}$ component, and $\beta$ denotes
the average transverse collective flow velocity
\cite{Khandai:2013fwa}.
In the high-$p_{\rm T}$ region,
Eq.~\ref{new_func_tsallis_distribution_function_second}
describes the power-law component modified by in-medium
energy loss. The parameter $A_{2}$ is the normalization
factor of the high-$p_{\rm T}$ component, $n_{2}$
determines the power-law behavior of the spectrum, and
$p_{2}$ represents the characteristic momentum scale of
this component. The parameter $\alpha$ characterizes
the dependence of the in-medium energy loss on the
energy of light quarks \cite{Baier:2000mf} \cite{De:2011fe}.
The parameter $B$ is associated with the characteristic size
of the medium, $q_{0}$ is fixed at 1~GeV/c, and $p_{2}$ is
not an independent fit parameter. The modified Tsallis
distribution has been shown to provide a good description
of particle spectra for $p_{\rm T}^{th} \gtrsim$~7~GeV/c
\cite{JPC_Kapil}.

In this work, the modified Tsallis parametrization described
by Eqs.~\ref{new_func_tsallis_distribution_function}
and~\ref{new_func_tsallis_distribution_function_second} is
applied to the $p_{\rm T}$ spectra of charged hadrons in $p+Pb$
and $Pb+Pb$ collisions. The fit parameters are extracted for
different centrality classes to investigate the medium-induced
modifications of the spectra. The resulting fit quality and
the centrality dependence of the extracted parameters are
discussed in the following section.
\FloatBarrier

\section{Results and discussion}
\subsection{$p+p$ collisions at $\sqrt{s}$ = 5.02 TeV}
\begin{figure}
\centering
\resizebox{0.5\textwidth}{0.3\textheight}{
\includegraphics{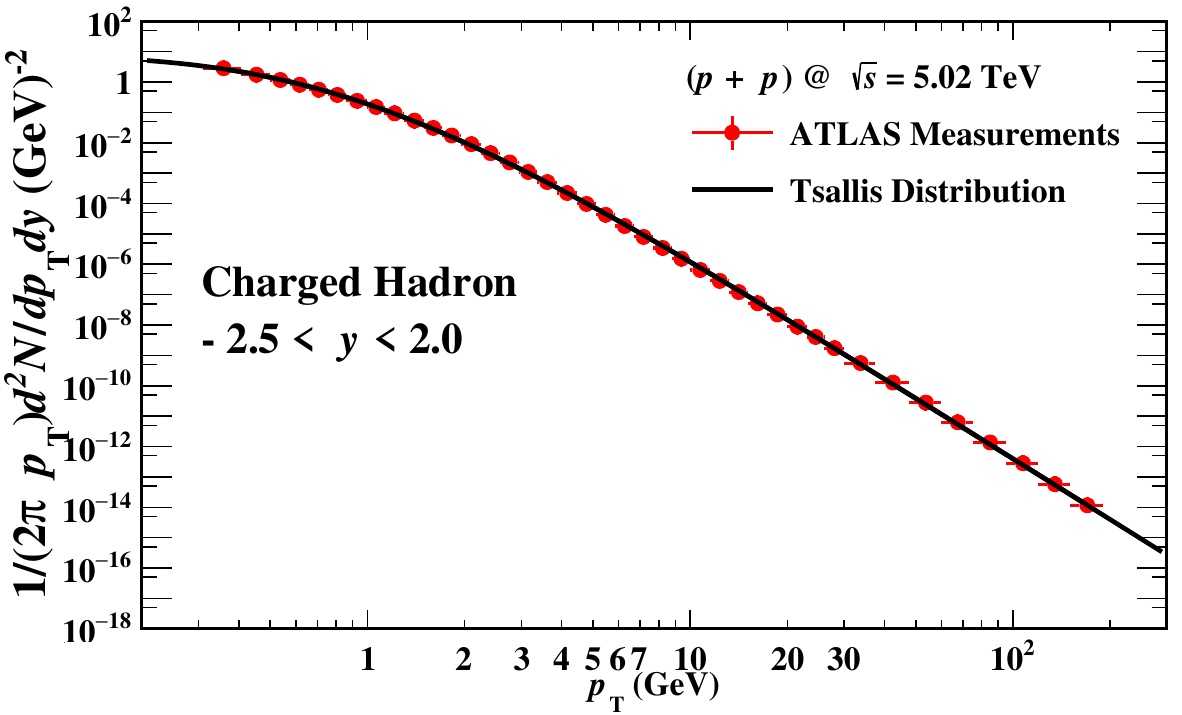}}
\caption{Invariant yields of charged hardons as a function of $p_{\rm{T}}$ 
in $p+p$ collisions at $\sqrt{s}~=$~5.02~TeV, measured by the ATLAS 
experiment \cite{ATLAS:2022kqu}. The solid curve represent fit obtained
using the Tsallis distribution function~(Eq.~\ref{Tsallis}).}
\label{Figure1_pp_collision_atlas}
\end{figure}

The invariant yields of charged hadrons as a function of $p_{\rm{T}}$ in
$p+p$ collisions at $\sqrt{s}$~=~5.02~TeV, measured by the ATLAS experiment 
\cite{ATLAS:2022kqu}, are described using the Tsallis distribution
(Eq.~\ref{Tsallis}). The corresponding spectra and the fitted distributions 
are shown in Fig.~\ref{Figure1_pp_collision_atlas}, where the solid curves
represent the best-fit results. The Tsallis distribution provides a good
description of the measured spectra, with the resulting
$\frac{\chi^{2}}{\rm NDF}$ values summarized in
Table~\ref{Table_one_pp_collision}.
\FloatBarrier

\subsection{$p+Pb$ collisions at $\sqrt{s_{\rm{NN}}}$ = 5.02 TeV}
\label{subsec1}
\begin{figure}
\centering
\resizebox{0.5\textwidth}{0.3\textheight}{
\includegraphics{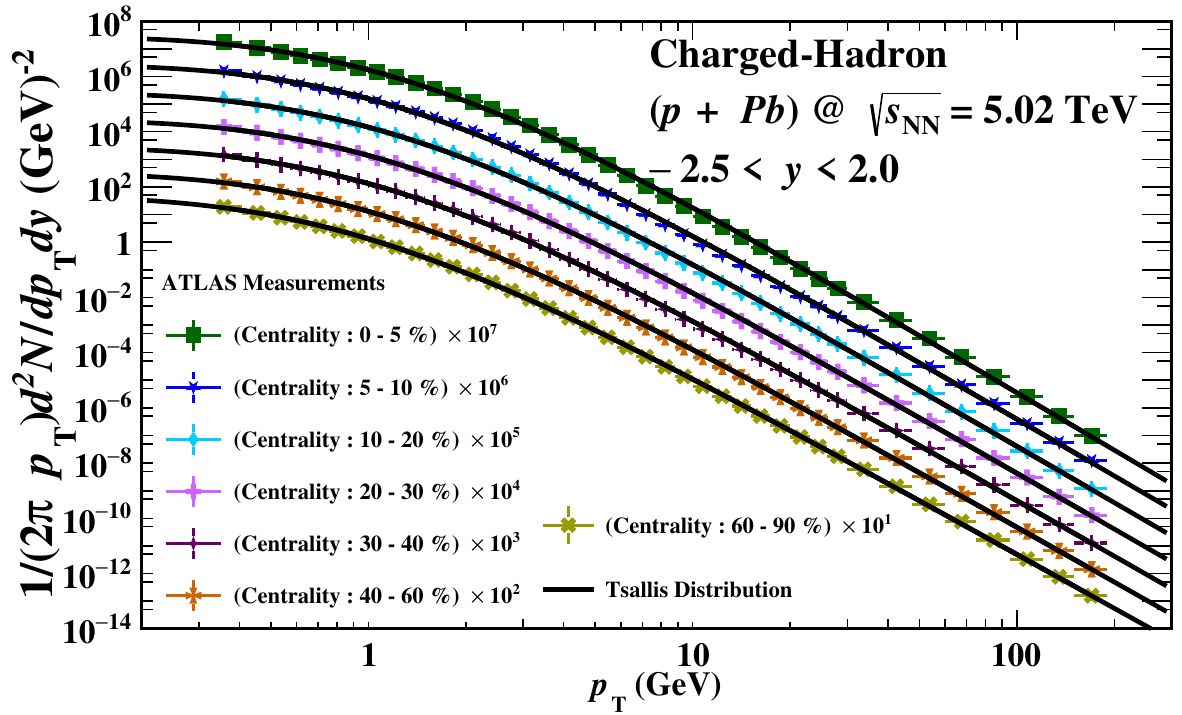}}
\caption{Invariant yields of charged hadrons as a function of $p_{\rm{T}}$
for different centrality classes in $p+Pb$ collisions at
$\sqrt{s_{\rm{NN}}}~=$~5.02 TeV, measured by the ATLAS Collaboration
\cite{ATLAS:2022kqu}. The solid curves represent Tsallis distribution fits
(Eq.~\ref{Tsallis}) to the data.}
\label{Figure2_pLead_502tev_tsallis}
\end{figure}

\begin{figure}
\centering
\resizebox{0.5\textwidth}{0.3\textheight}{
\includegraphics{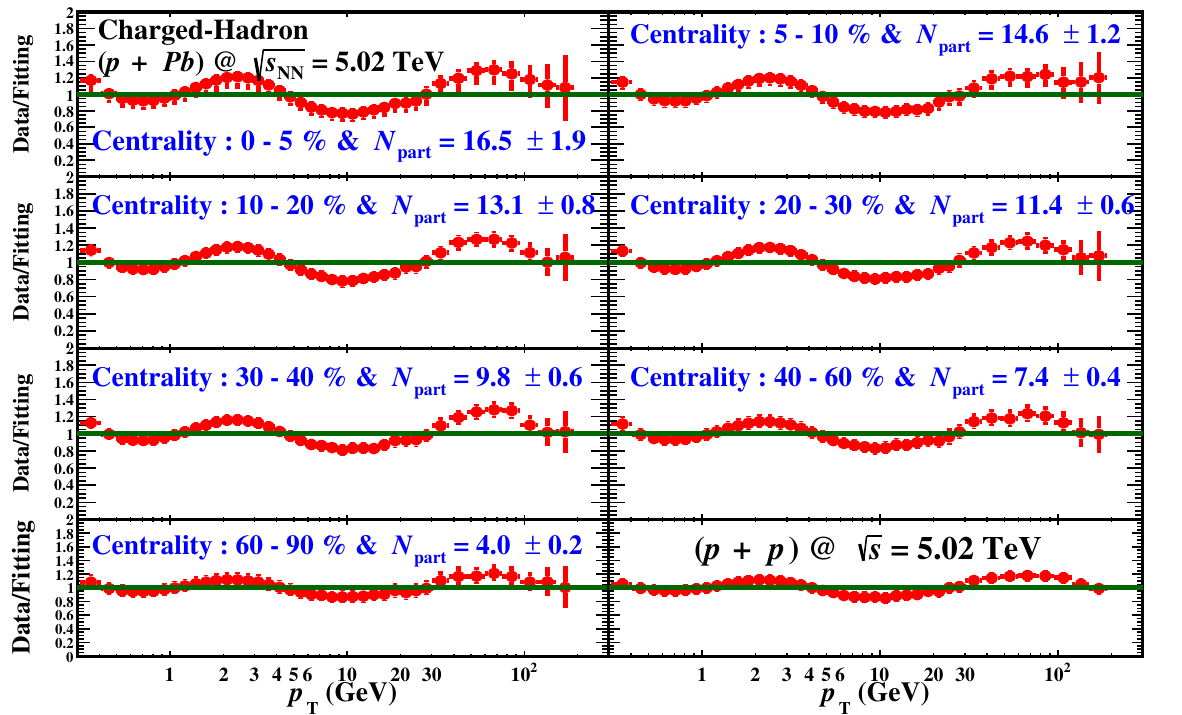}}
\caption{Ratio of the measured invariant yields of charged hadrons to the
Tsallis distribution fits (Eq.~\ref{Tsallis}) as a function of $p_{\rm{T}}$
for different centrality classes in $p+Pb$ collisions.
The observed deviations from unity quantify the limitations of the
conventional Tsallis parametrization in describing the measured spectra
over the full $p_{\rm T}$ range, motivating the use of the modified Tsallis
framework.}
\label{Figure3_pLead_502tev_databyfit}
\end{figure}

The $p_{\rm{T}}$ dependence of the invariant yields of charged hadrons for
different centrality classes in $p+Pb$ collisions at $\sqrt{s_{\rm{NN}}}~=$~
5.02~TeV is presented in Fig.~\ref{Figure2_pLead_502tev_tsallis} using
measurements from the ATLAS Collaboration~\cite{ATLAS:2022kqu}. The solid
curves represent fits using the Tsallis distribution~(Eq.~\ref{Tsallis}).
The comparison between the experimental data and the Tsallis distribution
fit~(Eq.~\ref{Tsallis}) is presented in
Fig.~\ref{Figure3_pLead_502tev_databyfit} through the data-to-fit ratio
as a function of $p_{\rm{T}}$ for different centrality classes in $p+Pb$
collisions at $\sqrt{s_{\rm{NN}}}~=$~5.02~TeV.
The Tsallis fit does not fully describe the measured $p+p$ and $p+Pb$
spectra, as indicated by the corresponding $\frac{\chi^{2}}{\rm NDF}$
values listed in Table~\ref{Table_two_pLead_collision_tsallis}.
The fitted parameters $n, q$, and $T$ for different centrality classes
and corresponding numbers of participants ($N_{\rm{part}}$) are also reported.
The results indicate that the fitted parameters $n$ and $q$ remain nearly
constant across all centrality classes, whereas $T$ exhibits a decreasing
trend with decreasing $N_{\rm{part}}$. This behavior suggests that the degree
of thermalization and collective medium effects gradually diminish from
central to peripheral collisions.


\begin{figure}
\centering
\resizebox{0.5\textwidth}{0.3\textheight}{
\includegraphics{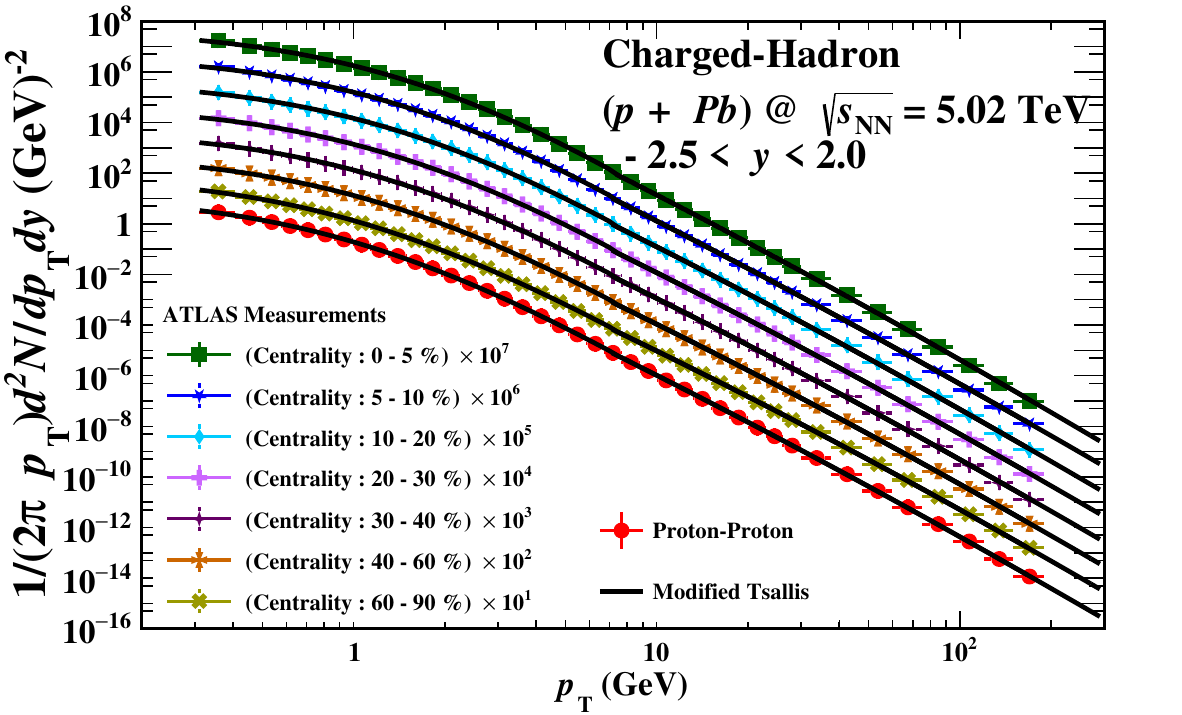}}
\caption{Invariant yields of charged hadrons as a function of $p_{\rm{T}}$  
for different centrality classes in $p+Pb$ collisions. The solid curves
represent the corresponding fits obtained using the modified Tsallis
parametrization given by
Eq.~\ref{modified_new_func_tsallis_distribution_function}.
The comparison illustrates the capability of the modified Tsallis
parametrization to describe the measured $p_{\rm{T}}$ spectra over the
investigated momentum range for different centrality intervals.
The agreement between the experimental data and the model provides
a basis for extracting the model parameters and investigating the
underlying medium-induced modifications in the production of
charged hadrons in $p+Pb$ collisions.}
\label{Figure4_pLead_502tev_tsallis_modified}
\end{figure}

\begin{figure}
\centering
\resizebox{0.5\textwidth}{0.3\textheight}{
\includegraphics{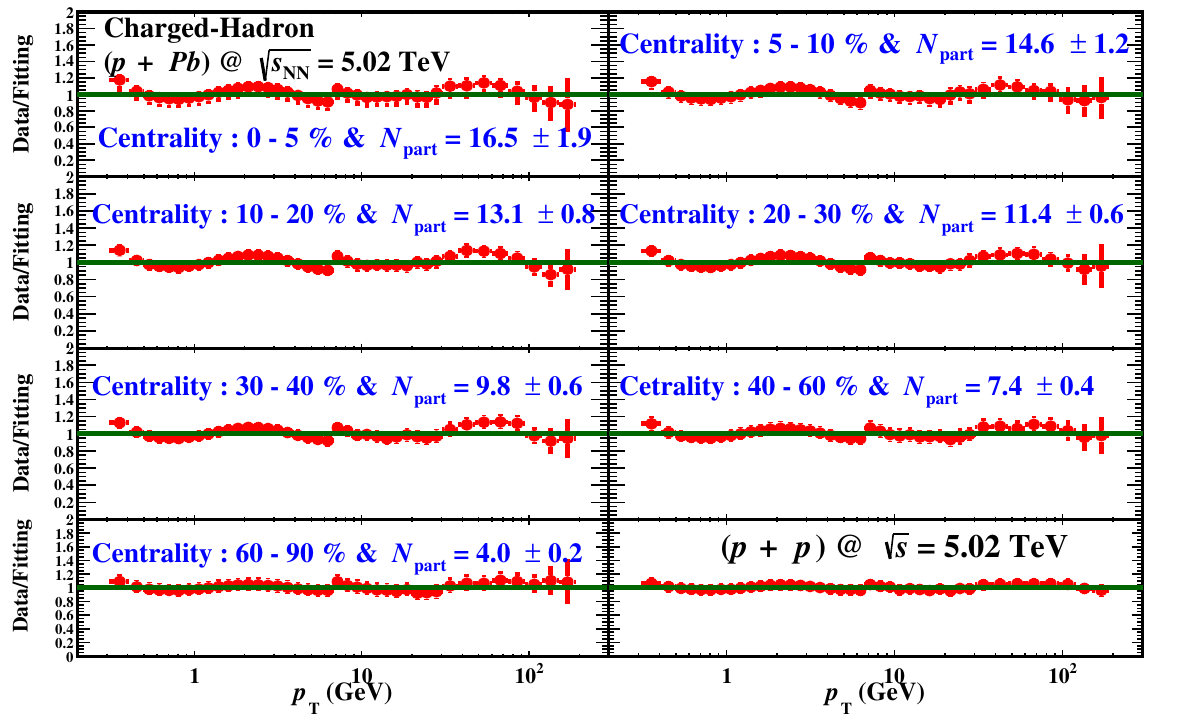}}
\caption{Ratio of the measured invariant yields of charged hadrons to the 
modified Tsallis parametrization given by
Eq.~\ref{modified_new_func_tsallis_distribution_function} as a function of
$p_{\rm{T}}$ for different centrality classes in $p+Pb$ collisions.
The ratio distributions provide a quantitative measure of the agreement
between the experimental spectra and the model description over the
investigated $p_{\rm{T}}$ range. The corresponding parameters of the modified
Tsallis functions extracted from the fits are summarized in
Table~\ref{Table_three_pLead_collision_tsallis_modified}.}
\label{Figure5_pLead_502tev_databyfit_modified}
\end{figure}

The deviations observed in the data-to-fit ratios in
Fig.~\ref{Figure3_pLead_502tev_databyfit} indicate the limitations of
the conventional Tsallis parametrization in describing the measured
spectra over the full $p_{\rm{T}}$ range. To investigate the
medium-induced modifications of charged-hadron production beyond the
conventional Tsallis description, the modified Tsallis parametrization
given by Eqs.~\ref{new_func_tsallis_distribution_function}
and~\ref{new_func_tsallis_distribution_function_second} is applied.
This formulation incorporates a low-$p_{\rm{T}}$ component associated
with transverse collective behavior and a high-$p_{\rm{T}}$ component
accounting for medium-induced partonic energy loss. The solid curves
in Fig.~\ref{Figure4_pLead_502tev_tsallis_modified} represent the
modified Tsallis fits to the measured spectra. The agreement between
the data and the parametrization is further evaluated through the
ratios of the measured spectra to the modified Tsallis function,
shown in Fig.~\ref{Figure5_pLead_502tev_databyfit_modified} as a
function of $p_{\rm{T}}$. The ratios demonstrate that the modified
Tsallis distribution provides a good description of the measured
spectra over the full $p_{\rm{T}}$ range for all centrality classes.
The extracted parameters of the modified Tsallis distribution for
different centrality classes and $N_{\rm{part}}$ are summarized in
Table~\ref{Table_three_pLead_collision_tsallis_modified}.
The parameters ($n_{1}, p_{1}, \beta$) associated with the
low-$p_{\rm{T}}$ component remain approximately constant across
different pseudorapidity intervals.
For the high-$p_{\rm{T}}$ component, the parameter $n_{2}$ is fixed
to 7.67 based on the value obtained from $p+p$ collisions, while
the exponent $\alpha$, which characterizes the energy dependence
of parton energy loss, remains unchanged. 
The extracted parameters provide insight into the underlying mechanisms
governing charged hadron production in $p+Pb$ collisions. Overall, the
modified Tsallis distribution described by
Eqs.~\ref{new_func_tsallis_distribution_function},
and~\ref{new_func_tsallis_distribution_function_second} provides a good
description of the charged hadron spectra over the full $p_{\rm{T}}$ range,
with the extracted parameters reflecting the underlying collective and
energy-loss mechanisms in $p+Pb$ collisions.

\FloatBarrier

\subsection{$Pb+Pb$ collisions at $\sqrt{s_{\rm{NN}}}$ = 5.02 TeV}
\label{subsec2}
\begin{figure}
\centering
\resizebox{0.5\textwidth}{0.3\textheight}{
\includegraphics{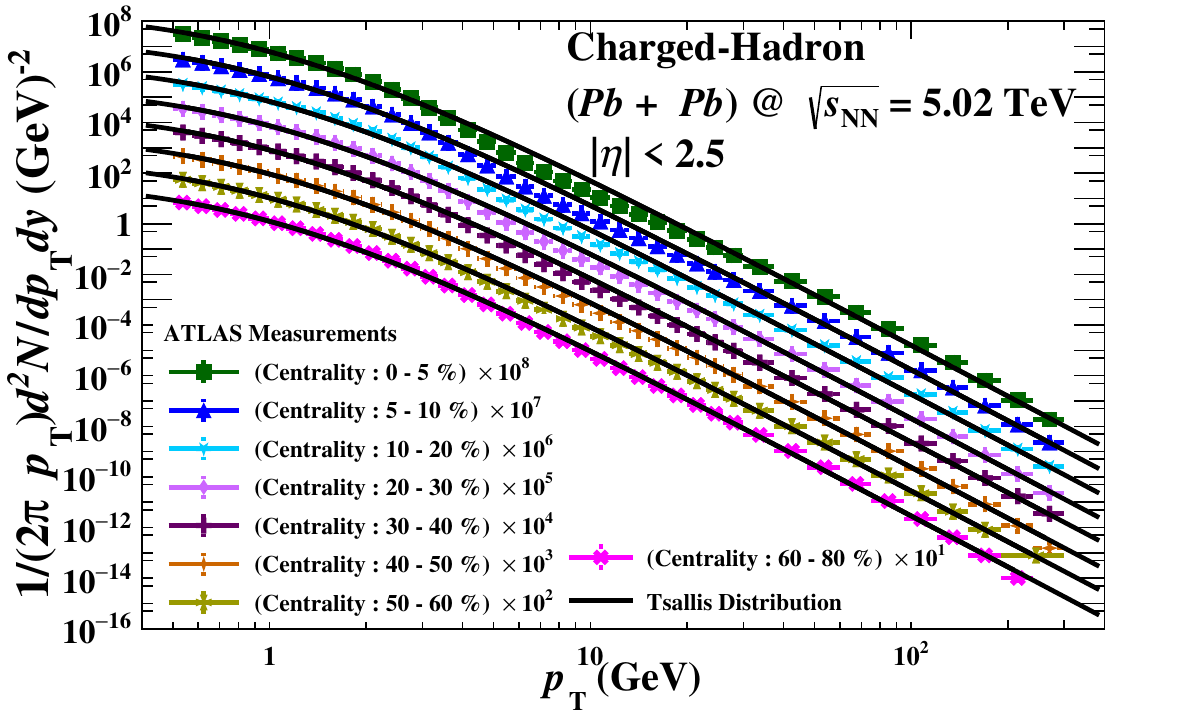}}
\caption{Invariant yields of charged hadrons as a function of $p_{\rm{T}}$ 
for different centrality classes in $Pb+Pb$ collisions at
$\sqrt{s_{\rm{NN}}}$ = 5.02 TeV, measured by the ATLAS Collaboration
\cite{ATLAS:2022kqu}. The solid curves demonstrate the results of fits
using the conventional Tsallis distribution function described by
Eq.~\ref{Tsallis}, which is employed to characterize the transverse
momentum dependence of the production of charged hadrons.}
\label{Figure6_LeadLead_502tev_tsallis}
\end{figure}

\begin{figure}
\centering
\resizebox{0.5\textwidth}{0.3\textheight}{
\includegraphics{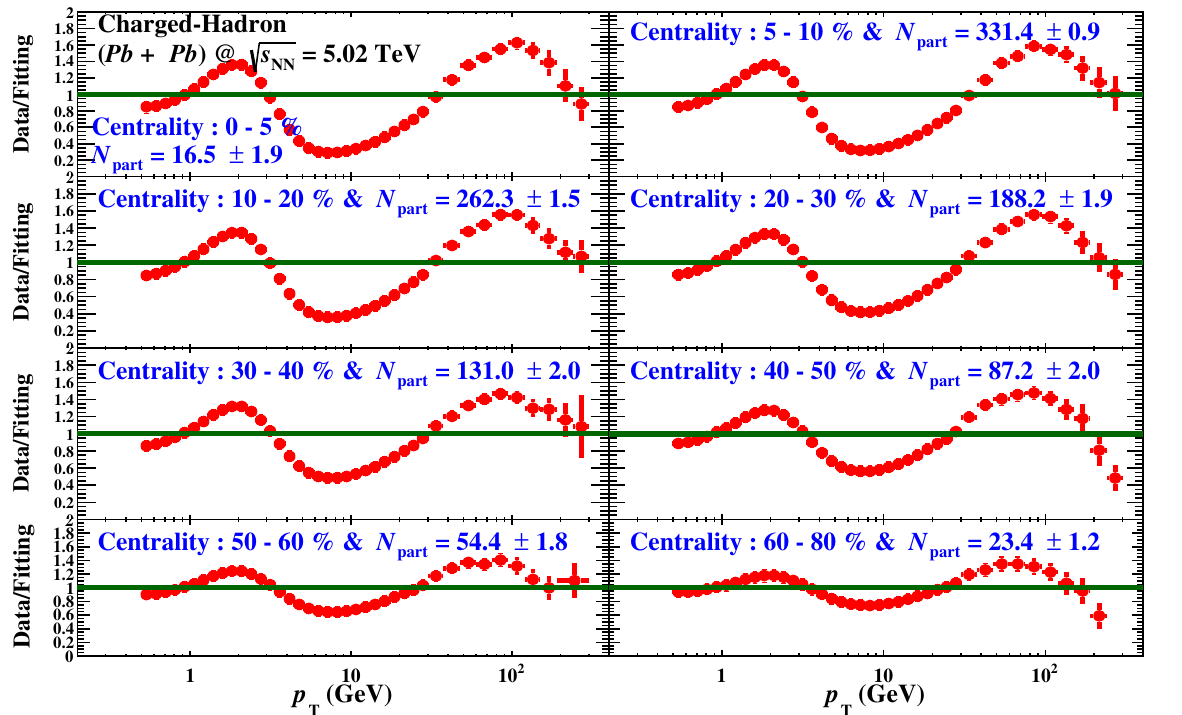}}
\caption{Ratio of the measured invariant yields of charged hadrons to the
fitted Tsallis distribution function described by Eq.~\ref{Tsallis} as a 
function of $p_{\rm{T}}$ for different centrality classes in $Pb+Pb$ collisions.
The ratio distributions demonstrate the level of agreement between the
experimental data and the conventional Tsallis parametrization, highlighting
the need for incorporating additional medium effects in the spectral
description.}
\label{Figure7_LeadLead_502tev_databyfit}
\end{figure}

The invariant yields of charged hadrons as a function of $p_{\rm{T}}$ for
different centrality classes in $Pb+Pb$ collisions at
$\sqrt{s_{\rm{NN}}}$ = 5.02 TeV, measured by the ATLAS Collaboration
\cite{ATLAS:2022kqu}, are presented in
Fig.~\ref{Figure6_LeadLead_502tev_tsallis}.
The solid curves represent the
corresponding fits obtained using the conventional Tsallis distribution
described by Eq.~\ref{Tsallis}. 
The ratio of the measured invariant yields to the Tsallis parametrization
is shown in Fig.~\ref{Figure7_LeadLead_502tev_databyfit} as a function of
$p_{\rm{T}}$ for the same collision system and centrality intervals.
The resulting $\frac{\chi^{2}}{\rm{NDF}}$ values, summarized in
Table~\ref{Table_five_pLead_collision_tsallis}, quantify the level
of agreement between the conventional Tsallis parametrization and the
experimental spectra.
The deviations observed in the data-to-fit ratios demonstrate that the
conventional Tsallis distribution alone does not provide an adequate
description over the full $p_{\rm{T}}$ range. These observations motivate
the use of the modified Tsallis framework, which incorporates
medium-induced effects in $Pb+Pb$ collisions.
\FloatBarrier

\begin{figure}
\centering
\resizebox{0.5\textwidth}{0.3\textheight}{
\includegraphics{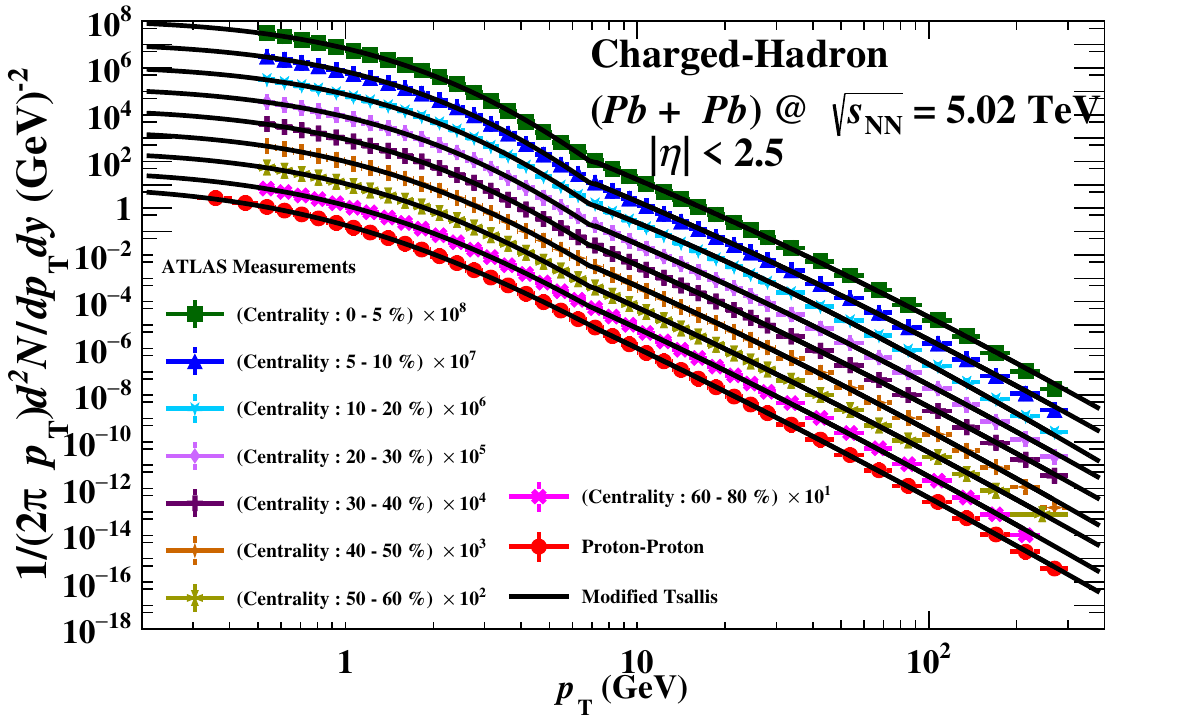}}
\caption{Invariant yields of charged hadrons as a function of $p_{\rm{T}}$
for different centrality classes in $Pb+Pb$ collisions. The solid curves
represent the corresponding fits obtained using the modified Tsallis
parametrization described by
Eq.~\ref{modified_new_func_tsallis_distribution_function}, which
incorporates the low- and high-$p_{\rm{T}}$ components of the particle
spectra and is used for the extraction of the model parameters.}
\label{Figure8_LeadLead_502tev_tsallis_modified}
\end{figure}

\begin{figure}
\centering
\resizebox{0.5\textwidth}{0.3\textheight}{
\includegraphics{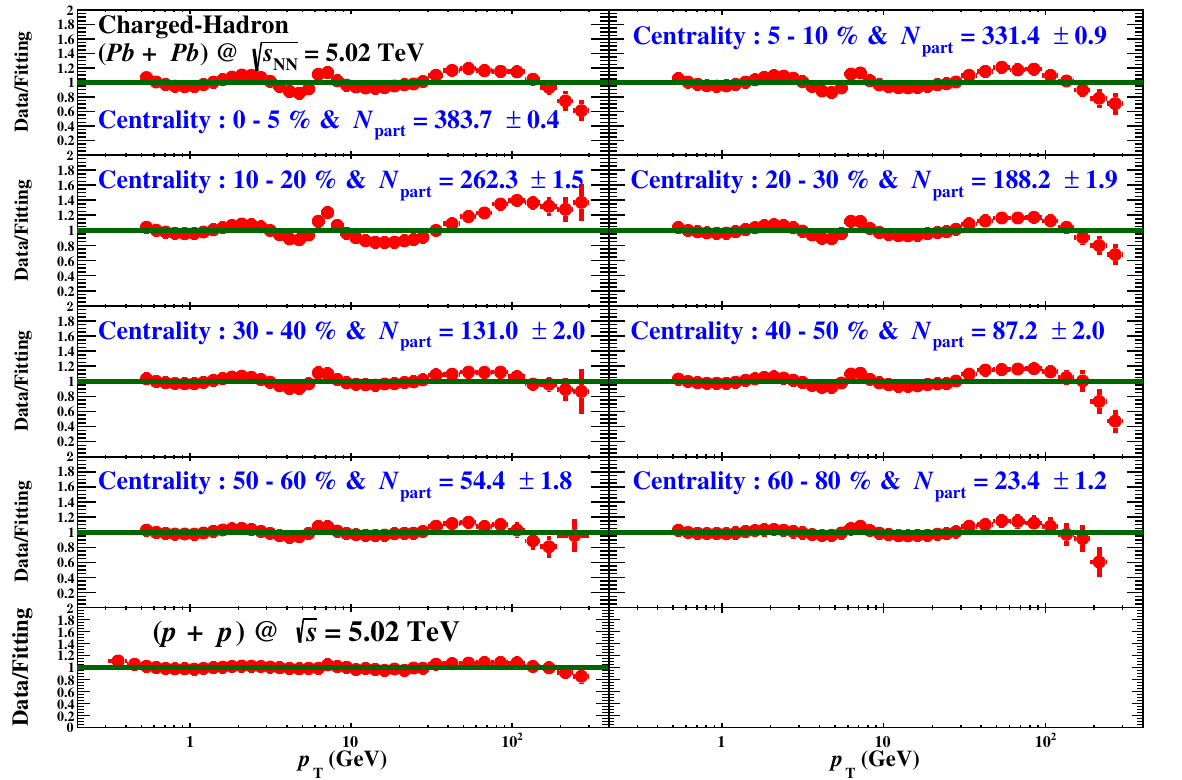}}
\caption{Relative comparison of the measured invariant yields of 
charged hadrons with the modified Tsallis distribution function
defined in Eq.~\ref{modified_new_func_tsallis_distribution_function}
for different centrality classes in $Pb+Pb$ collisions. The ratio
distributions provide a quantitative assessment of the consistency
between the experimental spectra and the modified Tsallis parametrization
over the entire investigated $p_{\rm{T}}$ range.}
\label{Figure9_LeadLead_502tev_databyfit_modified}
\end{figure}

The observed deviations between the measured spectra and the
conventional Tsallis parametrization, as quantified by the
ratio distributions in Fig.~\ref{Figure7_LeadLead_502tev_databyfit},
motivate the implementation of the modified Tsallis framework.
The parametrization given by
Eqs.~\ref{new_func_tsallis_distribution_function}
and~\ref{new_func_tsallis_distribution_function_second} incorporates the
low-$p_{\rm{T}}$ component associated with transverse collective flow and
the high-$p_{\rm{T}}$ component describing medium-induced partonic energy
loss. The resulting fits to the invariant yields of charged hadrons are
shown by the solid curves in
Fig.~\ref{Figure8_LeadLead_502tev_tsallis_modified}.
The ratio of the measured invariant yields to the modified Tsallis 
function is shown in Fig.~\ref{Figure9_LeadLead_502tev_databyfit_modified}
for different centrality intervals. The ratio distributions demonstrate
that the modified Tsallis framework provides a consistent description
of the measured spectra over the full $p_{\rm{T}}$ range for all centrality
classes.
The extracted parameters of the modified Tsallis distribution are
summarized in Table~\ref{Table_five_Pb_Pb_collisions_tsallis_modified}.
The parameters associated with the low-to-intermediate $p_{\rm{T}}$
component, $n_{1}$, $p_{1}$, and $\beta$, are taken to be constant across
the considered pseudo-rapidity ($|\eta|$) intervals.
For the high-$p_{\rm{T}}$ component, the parameter $n_{2}$ is fixed to 7.7,
motivated by the value obtained from $p+p$ collisions, while the exponent
$\alpha$, which characterizes the energy dependence of parton energy loss,
remains unchanged. The results demonstrate that the parametrization given
by Eqs.~\ref{new_func_tsallis_distribution_function}
and~\ref{new_func_tsallis_distribution_function_second} successfully
describes the $p_{\rm{T}}$ spectra of charged hadrons over a broad
$p_{\rm{T}}$ range. The extracted parameters provide quantitative insight
into the underlying mechanisms governing hadron production and
medium-induced modifications in $Pb+Pb$ collisions.
\FloatBarrier

\section{Conclusion}
We analyze the $p_{\rm{T}}$ spectra of charged hadrons for different
centrality classes in $p+Pb$ and $Pb+Pb$ collisions at
$\sqrt{s_{\rm{NN}}}$~=~5.02~TeV. The Tsallis distribution is first 
employed to describe the measured $p_{\rm{T}}$ spectra in both collision
systems. However, our analysis demonstrates that the conventional Tsallis
distribution alone does not provide an adequate description of the
experimental spectra over the entire $p_{\rm{T}}$ range. In particular,
a suppression of the charged hadrons yields is observed at high transverse
momentum ($p_{\rm{T}}>$~7~GeV/c), indicating the presence of medium-induced
modifications. To incorporate these effects, we employ a modified Tsallis
distribution that accounts for both collective behavior in the
low-to-intermediate $p_{\rm{T}}$ region and in-medium energy-loss effects
at high $p_{\rm{T}}$. The measured charged-hadron spectra for different
centrality classes in $p+Pb$ and $Pb+Pb$ collisions are then analyzed
using this modified parametrization.

The extracted parameters from the low-to-intermediate $p_{\rm{T}}$
region ($p_{\rm{T}}\leq$~7~GeV/c) provide evidence for transverse
collective flow effects. The parameters $n_{1}$, $p_{1}$, and
$\beta$ exhibit a clear centrality dependence, attaining larger
values for central collisions and gradually decreasing toward
peripheral collisions. This behavior can be attributed to the
increased number of multiple partonic scatterings in central
collisions, where the interaction density is significantly
higher compared with peripheral collisions. The observed 
centrality dependence therefore indicates the development of
transverse collective expansion among the produced particles.

At high transverse momentum ($p_{\rm{T}}>$~7~GeV/c), the exponent
$\alpha$, which characterizes the energy dependence of parton energy
loss, is extracted within the ranges 0.59$-$0.73 for $p+Pb$ collisions
and 0.32$-$0.59 for $Pb+Pb$ collisions. These values demonstrate the
significant influence of medium-induced energy loss on
high-$p_{\rm{T}}$ partons. Overall, the modified Tsallis distribution
successfully describes the charged hadrons transverse momentum spectra
in both collision systems and enables a quantitative characterization
of the underlying medium effects. The extracted parameters establish
a direct phenomenological connection between the spectral modifications
and the dynamical properties of the produced medium, including
transverse collective flow and in-medium parton energy loss in
relativistic heavy-ion collisions.

\section*{Acknowledgement}
The author Venktesh Singh acknowledges financial support from the
DST, DST-FIST and DST-PURSE.
The author P. K. Khandai acknowledges financial support from the
University Grants Commission through the Start-Up Research Grant project.
\section*{Data}
The data analyzed in this study are publicly available through the ATLAS
Collaboration and are also available from the authors upon reasonable
request.
\FloatBarrier

\section*{Appendix}
\begin{table}[ht]
\caption{Parameters of the Tsallis distribution (Eq.~\ref{Tsallis}) extracted  
from fit to the charged hadron $p_{\rm{T}}$ spectra in $p+p$ collisions at 
$\sqrt{s}~=$~5.02~TeV.}
\label{Table_one_pp_collision}
\begin{center}
\begin{tabular}{|c || c | c | c | c | c} \hline
 Rapidity   &  $n$ & $q$ &   $T$  & $\frac{\chi^{2}}{\rm{NDF}}$ \\ 
 ($y$)   &      &     & (MeV)  &               \\ \hline \hline
 - 2.5 $< y <$ 2.0 & 7.67 $\pm$ 0.10 & 1.13 & 84.78 $\pm$ 5.46 & 0.05 \\ \hline 
\end{tabular}
\end{center}
\end{table}

\begin{table}[ht]
\caption{Parameters of the Tsallis distribution function 
(Eq.~\ref{Tsallis}) obtained from fits to the $p_{\rm{T}}$ spectra of
charged hadrons for different centrality classes in $p+Pb$ collisions at
$\sqrt{s_{\rm{NN}}} = 5.02$ TeV. The corresponding values of
$\chi^{2}/\rm{NDF}$ are also listed.}
\label{Table_two_pLead_collision_tsallis}
\begin{center}
\scalebox{0.8}{
\begin{tabular}{| c || c | c | c | c | c | c |} \hline
Centrality   & $N_{\rm{part}}$ & $n$ & $q$ &  $T$  & $\frac{\chi^{2}}{\rm{NDF}}$ \\ 
($\%$)       &               &     &     & (MeV) &   \\ \hline \hline
0  - 5   & 16.50 $\pm$ 1.90  & 7.97 $\pm$ 0.11  & 1.13  & 116.26 $\pm$ 5.98  & 0.16 \\  
5  - 10  & 14.60 $\pm$ 1.20  & 7.88 $\pm$ 0.11  & 1.13  & 112.85 $\pm$ 6.14  & 0.15 \\ 
10 - 20  & 13.10 $\pm$ 0.80  & 7.83 $\pm$ 0.10  & 1.13  & 110.56 $\pm$ 5.40  & 0.14 \\  
20 - 30  & 11.40 $\pm$ 0.60  & 7.77 $\pm$ 0.10  & 1.13  & 107.15 $\pm$ 5.54  & 0.13 \\  
30 - 40  & 9.80  $\pm$ 0.60  & 7.73 $\pm$ 0.10  & 1.13  & 104.12 $\pm$ 5.56  & 0.11 \\  
40 - 60  & 7.40  $\pm$ 0.40  & 7.65 $\pm$ 0.10  & 1.13  & 98.44  $\pm$ 5.70  & 0.09 \\  
60 - 90  & 4.00  $\pm$ 0.20  & 7.53 $\pm$ 0.10  & 1.13  & 86.65  $\pm$ 5.77  & 0.05 \\ \hline
\end{tabular}}
\end{center}
\end{table}

\begin{table}[ht]
\caption{Parameters of the modified Tsallis parametrization described by
Eqs.~\ref{new_func_tsallis_distribution_function},
and~\ref{new_func_tsallis_distribution_function_second}, obtained from
fits to the $p_{\rm{T}}$ spectra of charged hadrons for different centrality
classes in $p+Pb$ collisions. The corresponding fit parameters characterize
the low- and high-$p_{\rm{T}}$ components of the production spectra of charged
hadrons, and the resulting $\frac{\chi^{2}}{\rm NDF}$ values are also listed.}
\label{Table_three_pLead_collision_tsallis_modified}
\begin{center}
\begin{tabular}{|c || c | c | c | c | c | c | c | c |}  \hline \hline 
System & Centrality & $N_{\rm{part}}$ & $n_{1}$ & $p_{1}$    & $\beta$ &  $\alpha$ & $B$         & $\frac{\chi^{2}}{\rm{NDF}}$  \\
&   ($\%$)  &                &         & (GeV/$c$) &         &            & (GeV/$c$)  & \\ \hline
$p+Pb$    &   0 - 5            & 16.50 $\pm$ 1.90 & 7.69 $\pm$ 2.21 & 1.38 $\pm$ 1.55 & 0.13 $\pm$ 0.55 & 0.59 $\pm$ 0.27 & 2.40 $\pm$ 0.75 & 0.05 \\ 
          
$p+Pb$    &   5 - 10           & 14.60 $\pm$ 1.20 & 7.36 $\pm$ 0.84 & 1.23 $\pm$ 0.28 & 0.15 $\pm$ 0.30 & 0.63 $\pm$ 0.25 & 2.76 $\pm$ 1.35 & 0.05 \\ 
          
$p+Pb$    &   10 - 20           & 13.10 $\pm$ 0.80 & 7.32 $\pm$ 1.88 & 1.22 $\pm$ 0.79 & 0.13 $\pm$ 0.12 & 0.59 $\pm$ 0.13 & 2.72 $\pm$ 0.47 & 0.05 \\  
          
$p+Pb$    &   20 - 30           & 11.40 $\pm$ 0.60 & 7.20 $\pm$ 2.19 & 1.16 $\pm$ 0.81 & 0.13 $\pm$ 0.19 & 0.60 $\pm$ 0.05 & 2.90 $\pm$ 0.38 & 0.04 \\ 
          
$p+Pb$    &   30 - 40           & 9.80 $\pm$ 0.60 & 7.12 $\pm$ 1.96 & 1.11 $\pm$ 0.68 & 0.13 $\pm$ 0.10 & 0.60 $\pm$ 0.05 & 2.95 $\pm$ 0.38 & 0.04 \\  
          
$p+Pb$    &   40 - 60           & 7.40 $\pm$ 0.40 & 6.99 $\pm$ 2.03 & 1.02 $\pm$ 0.64 & 0.13 $\pm$ 0.09 & 0.59 $\pm$ 0.05 & 3.08 $\pm$ 0.39 & 0.03 \\  
          
$p+Pb$    &   60 - 90           & 4.00 $\pm$ 0.20 & 6.78 $\pm$ 2.30 & 0.86 $\pm$ 0.45 & 0.12 $\pm$ 0.10 & 0.57 $\pm$ 0.04 & 3.26 $\pm$ 0.39 & 0.02 \\  
          
$p+p$    &    -                &  -  & 6.94 $\pm$ 2.36 & 0.86 $\pm$ 1.17 & 0.12 $\pm$ 0.10 & 0.61 $\pm$ 0.05 & 2.92 $\pm$ 0.37 & 0.01 \\ \hline
\end{tabular}
\end{center}
\end{table}

\begin{table}[ht]
\caption{Parameters of the conventional Tsallis distribution function 
given in Eq.~\ref{Tsallis}, extracted from fits to the $p_{\rm{T}}$ spectra
of charged hadrons for different centrality classes in $Pb+Pb$ collisions.
The associated $\frac{\chi^{2}}{\rm{NDF}}$ values are provided to evaluate
the performance of the standard Tsallis parametrization in describing the
measured spectra.}
\label{Table_five_pLead_collision_tsallis}
\begin{center}
\begin{tabular}{|c || c | c | c | c | c | c |} \hline
Centrality   & $N_{\rm{part}}$ & $n$ & $q$ &  $T$  & $\frac{\chi^{2}}{\rm{NDF}}$ \\ 
($\%$)       &               &     &     & (MeV) &   \\ \hline \hline 
0 - 5   & 383.70 $\pm$ 0.40 & 7.71 $\pm$ 0.07 & 1.13 & 94.07 $\pm$ 5.22  & 1.05 \\
5 - 10  & 331.40 $\pm$ 0.90 & 7.70 $\pm$ 0.10 & 1.13 & 95.28 $\pm$ 8.90  & 0.98 \\ 
10 - 20 & 262.30 $\pm$ 1.50 & 7.70 $\pm$ 0.12 & 1.13 & 95.85 $\pm$ 11.09 & 0.87 \\
20 - 30 & 188.20 $\pm$ 1.90 & 7.71 $\pm$ 0.10 & 1.13 & 96.33 $\pm$ 8.36  & 0.73 \\  
30 - 40 & 131.00 $\pm$ 2.00 & 7.67 $\pm$ 0.09 & 1.13 & 94.27 $\pm$ 7.99  & 0.58 \\ 
40 - 50 & 87.20  $\pm$ 2.00 & 7.71 $\pm$ 0.10 & 1.13 & 95.12 $\pm$ 7.83  & 0.43 \\  
50 - 60 & 54.40  $\pm$ 1.80 & 7.68 $\pm$ 0.11 & 1.13 & 92.76 $\pm$ 7.63  & 0.29 \\ 
60 - 80 & 23.40  $\pm$ 1.20 & 7.72 $\pm$ 0.11 & 1.13 & 92.51 $\pm$ 7.39  & 0.16 \\ \hline
\end{tabular}
\end{center}
\end{table}

\clearpage
\begin{table}[h!] 
\caption{Summary of the parameters extracted from the modified Tsallis 
fits to the $p_{\rm{T}}$ spectra of charged hadrons in $Pb+Pb$ collisions.
The parametrization used in the fits is described
by Eqs.~\ref{new_func_tsallis_distribution_function},
and~\ref{new_func_tsallis_distribution_function_second}, and the
corresponding $\frac{\chi^{2}}{\rm{NDF}}$ values are reported as a measure
of the fit quality.}
\label{Table_five_Pb_Pb_collisions_tsallis_modified}
\begin{center}
\begin{tabular}{|c || c | c | c | c | c | c | c | c |}  \hline \hline
System & Centrality & $N_{\rm{part}}$ & $n_{1}$ & $p_{1}$    & $\beta$ &  $\alpha$ & $B$         & $\frac{\chi^{2}}{\rm{NDF}}$  \\
&   ($\%$)  &                &         & (GeV/$c$) &         &            & (GeV/$c$)  & \\ \hline
$Pb+Pb$  & 0 - 5 & 383.70 $\pm$ 0.40 & 10.85 $\pm$ 0.43 & 2.00 $\pm$ 1.77 & 0.28 $\pm$ 0.10 & 0.49 $\pm$ 0.05 & 4.97 $\pm$ 0.57 & 0.06 \\ 

$Pb+Pb$    &   5 - 10           & 331.40 $\pm$ 0.90 & 10.72 $\pm$ 0.44 & 2.00 $\pm$ 1.06 & 0.27 $\pm$ 0.10 & 0.50 $\pm$ 0.13 & 4.83 $\pm$ 0.71 & 1.29 \\  

$Pb+Pb$    &   10 - 20          & 262.30 $\pm$ 1.50 & 10.56 $\pm$ 0.47 & 2.00 $\pm$ 1.03 & 0.25 $\pm$ 0.11 & 0.32 $\pm$ 0.06 & 4.61 $\pm$ 0.67 & 1.28 \\ 
       
$Pb+Pb$    &   20 - 30          & 188.20 $\pm$ 1.90 & 10.35 $\pm$ 0.72 & 2.00 $\pm$ 1.23 & 0.21 $\pm$ 0.12 & 0.50 $\pm$ 0.14 & 4.26 $\pm$ 0.60 & 0.07 \\  

$Pb+Pb$    &   30 - 40          & 131.00 $\pm$ 2.00 & 10.09 $\pm$ 0.96 & 2.00 $\pm$ 1.41 & 0.15 $\pm$ 0.09 & 0.52 $\pm$ 0.12 & 4.02 $\pm$ 0.55 & 0.02 \\ 

$Pb+Pb$    &   40 - 50          & 87.20 $\pm$ 2.00 & 8.61 $\pm$ 4.54 & 1.43 $\pm$ 1.27 & 0.24 $\pm$ 0.13 & 0.48 $\pm$ 0.17 & 3.61 $\pm$ 0.49 & 1.16 \\ 

$Pb+Pb$    &   50 - 60          & 54.40 $\pm$ 1.80 & 7.96 $\pm$ 4.10 & 1.22 $\pm$ 0.83 & 0.24 $\pm$ 0.17 & 0.55 $\pm$ 0.16 & 3.56 $\pm$ 0.77 & 0.02 \\  

$Pb+Pb$    &   60 - 80          & 23.40 $\pm$ 1.20 & 7.60 $\pm$ 2.35 & 1.11 $\pm$ 0.85 & 0.19 $\pm$ 0.13 & 0.52 $\pm$ 0.21 & 3.01 $\pm$ 0.52 & 0.02 \\ 
          
$p+p$    &     -                &  -  & 7.45 $\pm$ 0.52 & 1.11 $\pm$ 0.74 & 0.02 $\pm$ 0.01 & 0.59 $\pm$ 0.12 & 2.86 $\pm$ 0.65 & 0.01 \\ \hline
\end{tabular}
\end{center}
\end{table}

\clearpage


\begin{thebibliography}{}
\bibitem{RHIC1}
K. Adcox (PHENIX Collaboration),
Nucl. Phys. A {\textbf 757}, (2005) 184-283.
\bibitem{RHIC2}
J. Adams (STAR Collaboration),
Nucl. Phys. A {\textbf 757}, (2005) 102-183.
\bibitem{LHC}
N. Armesto (LHC Collaboration), J. Phys. G {\textbf 35}, (2008) 054001.
\bibitem{Shuryak}
E. V. Shuryak, Phys. Rept. {\textbf 61}, (1980) 71-158.
\bibitem{Olive}
K. A. Olive, Science {\textbf 251}, (1991) 1194-1199.
\bibitem{Schwarz}
D. J. Schwarz, Annalen Phys {\textbf 12}, (2003) 220-270.
\bibitem{PPPROD}
F. Becattini and U. Heinz, Z. Phys. C{\textbf 76}, (1997) 269.
\bibitem{HighMul}
V. Khachatryan (CMS Collaboration), JHEP {\textbf 09}, (2010) 091.
\bibitem{HighMul2}
J. Adam (ALICE Collaboration),
Nature Phys. {\textbf 13}, (2017) 535.
\bibitem{Cleymans:1985wb}
J.~Cleymans, R.~V.~Gavai and E.~Suhonen,
Phys. Rept. \textbf{130}, (1986) 217.
\bibitem{Singh:1992sp}
C.~P.~Singh, Phys. Rept. \textbf{236}, (1993) 147-224.
\bibitem{Jaiswal:2020hvk}
A.~Jaiswal, N.~Haque, A.~Abhishek, R.~Abir, A.~Bandyopadhyay, K.~Banu,
S.~Bhadury, S.~Bhattacharyya, T.~Bhattacharyya and D.~Biswas,
\textit{et al.}, Int. J. Mod. Phys. E \textbf{30}, (2021) 2130001.
\bibitem{Saraswat:2022zcn}
K.~Saraswat, D.~S.~Rawat and H.~C.~Chandola,
Nucl. Phys. A \textbf{1022}, (2022) 122441.
\bibitem{Cho:1979nv}
Y.~M.~Cho,  Phys. Rev. D \textbf{21}, (1980) 1080.
\bibitem{Cho:2002iv}
Y.~M.~Cho and D.~G.~Pak, Phys. Rev. D \textbf{65}, (2002) 074027.
\bibitem{Chandola:2009zz}
H.~C.~Chandola and D.~Yadav, Nucl. Phys. A \textbf{829}, (2009) 151-169.
\bibitem{Collective_flow}
T. Hirano and Y. Nara, Phys. Rev. C {\textbf 69}, (2004) 034908.
\bibitem{Jet_Quenching}
X. N. Wang, Phys. Lett. B {\textbf 579}, (2004) 299.
\bibitem{Tsallis:1987eu}
C.~Tsallis, J. Statist. Phys. {\textbf 52}, (1988) 479.
\bibitem{Biro:2008hz}
T. S. Biro, G. Purcsel and K. Urmossy,
Eur. Phys. J. A {\textbf 40}, (2009) 325.
\bibitem{PPG099}
A. Adare (PHENIX Collaboration),
Phys.Rev. D {\bf 83}, (2011) 052004.
\bibitem{q_Tsallis}
G. Wilk and Z. Wlodarczyk,
Phys. Rev. Lett. {\bf 84}, (2000) 2770.
\bibitem{IJMPA_Khandai}
P. K. Khandai, P. Sett, P. Shukla and V. Singh,
Int. J. Mod. Phys. A {\bf 28}, (2013) 1350066.
\bibitem{Wong:2012zr}
C. Y. Wong and G. Wilk,
Acta Phys. Polon. B {\bf 43}, (2012) 2047-2054.
\bibitem{Wong:2013sca}
C. Y. Wong and G. Wilk,
Phys. Rev. D {\bf 87}, (2013) 114007.
\bibitem{Hagedorn:1983wk}
R. Hagedorn, Riv. Nuovo Cim. {\bf 6N10}, (1983).
\bibitem{Blankenbecler:1974tm}
R. Blankenbecler and S. J. Brodsky,
Phys. Rev. D {\bf 10}, (1974) 2973.
\bibitem{JPC_Kapil}
K. Saraswat, P. Shukla and V. Singh, 
J. Phys. Commun. {\bf 2}, (2018) 035003.
\bibitem{Khandai}
P. Kumar, P. K. Khandai, Kapil Saraswat and V. Singh,
Int. J. Mod. Phys. A {\bf 36}, (2021) 2150059.
\bibitem{ATLAS:2022kqu}
G. Aad (ATLAS collaboration),
J. High Energ. Phys. {\bf 07}, (2013) 74.
\bibitem{Tang:2008ud}
Z.~Tang, Y.~Xu, L.~Ruan, G.~van Buren, F.~Wang and Z.~Xu,
Phys.\ Rev.\ C {\bf 79}, (2009) 051901.
\bibitem{Khandai:2013fwa}
P.~K.~Khandai, P.~Sett, P.~Shukla and V.~Singh,
J.\ Phys.\ G {\bf 41}, (2014) 025105.
\bibitem{Sett:2015lja}
P.~Sett and P.~Shukla,
Int.\ J.\ Mod.\ Phys.\ E {\bf 24}, (2015) 1550046.
\bibitem{Baier:2000mf}
R.~Baier, D.~Schiff and B.~G.~Zakharov,
Ann.\ Rev.\ Nucl.\ Part.\ Sci. {\bf 50}, (2000) 37-69.
\bibitem{De:2011fe}
S.~De and D.~K.~Srivastava, J.\ Phys.\ G {\bf 39}, (2012) 015001. 
Erratum : [J.\ Phys.\ G] {\bf 40}, (2013) 049502. 
\bibitem{root:usersguide:fitting} ROOT User's Guide : Fitting Histograms,
CERN ROOT, 2023.
\{https://root.cern/root/htmldoc/guides/users-guide/FittingHistograms.html\}.
\bibitem{minuit2:guide} MINUIT2 $-$ Function Minimization in ROOT, 2023.
\{https://root.cern/root/htmldoc/guides/minuit2/Minuit2.html\}.
\bibitem{bak::cou}
S.~Baker and R.D.~Cousins, 
Nucl. Instrum. Meth. Phys. Res. {\bf 221}, (1984) 437.
\end{thebibliography}
\end{document}